\documentclass[12pt, preprint]{aastex}

\newcommand{\jpb}{J. Phys. B}
\newcommand{\hetgs}{HETGS}
\newcommand{\ngc}{NGC}
\newcommand{\kms}{km\,s$^{-1}$}
\begin{document}

\title{Inner-shell Absorption Lines of Fe~VI-- Fe~XVI: A Many-body
  Perturbation Theory Approach}
\author{
Ming F. Gu\altaffilmark{1},
Tomer Holczer\altaffilmark{2},
Ehud Behar\altaffilmark{2}, and
Steven M. Kahn\altaffilmark{1}
}
\altaffiltext{1}{Kavli Institute for Particle Astrophysics and Cosmology
and Department of Physics, Stanford University, CA 94305, USA}
\altaffiltext{2}{Department of Physics, Technion, Haifa 32000, Israel}

\begin{abstract}
We provide improved atomic calculation of wavelengths, oscillator
strengths, and autoionization rates relevant to the $2\to 3$
inner-shell transitions of Fe~VI--XVI, the so-called Fe~M-shell
unresolved transition array (UTA). A second order many-body
perturbation theory is employed to obtain accurate transition
wavelengths, which are systematically larger than previous
theoretical results by 15--45~m{\AA}. For a few transitions of
Fe~XVI and Fe~XV where laboratory measurements exist, our new
wavelengths are accurate to within a few m{\AA}. Using these new
calculations, the apparent discrepancy in the velocities between
the Fe~M-shell UTA and other highly ionized absorption lines in
the outflow of NGC~3783 disappears. The oscillator strengths in
our new calculation agree well with the previous theoretical data,
while the new autoionization rates are significantly larger,
especially for lower charge states. We attribute this discrepancy
to the missing autoionization channels in the previous
calculation. The increased autoionization rates may slightly
affect the column density analysis of the Fe~M-shell UTA for
sources with high column density and very low turbulent
broadening. The complete set of atomic data is provided as an
electronic table.
\end{abstract}


\section{Introduction}
Since the first astrophysical detection of $2\to 3$ innershell
absorption lines of Fe~VII--XII in the X-ray spectrum of IRAS
13349+2438 obtained with the Reflection Grating Spectrometer (RGS)
on board \textit{XMM-Newton} \citep{sako01}, this so-called
Fe~M-shell unresolved transition array (UTA) has been identified
in many soft X-ray sources. The UTA is mainly comprised of a
cluster of lines originating from $2p$--$3d$ transitions of
M-shell iron ions that are located between 15--17~{\AA}. These
lines, when properly modeled, provide important information on the
ionization structure, column density, and outflow kinematics of
the absorbing materials. \citet{behar01} calculated a complete set
of atomic data for such modeling using the Hebrew University
Lawrence Livermore Atomic Code \citep[HULLAC,]{barshalom01}, and provided an
abbreviated list of transition
wavelengths, oscillator strengths, and autoionization rates. That
complete data set has also been incorporated in some commonly used
plasma modeling codes.

In most observations where the Fe~M-shell UTA has been identified,
individual features from different ionization states are not
resolved, mostly due to significant turbulent velocity broadening
($>700$~\kms) at the source. Low statistics and insufficient
spectral resolution may also hamper the identification of
individual charge states. Therefore, one has to depend on global
photoionized plasma models to constrain the physical conditions of
the absorbing gas. In global fits, where individual UTAs are
unresolved, the exact rest-frame position of each UTA is
particularly important, especially since current photoionization
balance calculations fail to predict the formation of Fe~M-shell
ions consistently with highly ionized species of other elements
\citep{netzer03,netzer04,kraemer04}.

Recently, \citet{holczer05} have reexamined the \textit{Chandra}
High Energy Transmission Grating Spectrometer (HETGS) spectrum of
the active galactic nucleus (AGN) of NGC~3783, by coadding all
900~ks available observations. The low turbulent velocity of the
outflow of NGC~3783, the high spectral resolution of the HETGS,
and the very long exposure time enabled the individual absorption
features from different Fe~M-shell ions to be resolved. The
analysis of \citet{holczer05} showed that the outflow velocity
associated with the Fe~M-shell UTA appears to be different from
those of other highly-ionized species, namely, Fe~XVII, O~VII, and
O~VIII. In fact, the data prefer a zero velocity for the gas
associated with Fe~M-shell ions, and a $-590$~km\,s$^{-1}$ outflow
velocity for the gas associated with more highly ionized ions.
Therefore, a two component model was proposed where the Fe~M-shell
ions are not part of the AGN outflow, with an important
implication of much lower mass loss rate.

As stated in \citet{holczer05}, the conclusions of that paper
depend strongly on the calculated HULLAC wavelengths. The
velocities of the Fe~M-shell UTA can be distinguished from those
of other ions only if the HULLAC wavelengths are accurate to better than
30~m{\AA}, which corresponds to a velocity of $\sim$590~\kms\ at 16~m{\AA}.
However, it has been shown that the theoretical method employed by
HULLAC can generally produce uncertainties of up to 20--50~m{\AA}
for lines in the 10--20~{\AA} band \citep{brown02, gu05}.
Unfortunately, no laboratory measurements exist for the
wavelengths of relevant ions except for a few strong lines of
Fe~XVI and Fe~XV \citep{brown01}. For these few lines, the HULLAC
wavelengths are indeed underestimated by 20--30~m{\AA}. However,
the strongest and cleanest absorption features in the observed
spectrum of NGC~3783 are from Fe IX--XI. If the HULLAC wavelengths
for these ions are also underestimated by similar amounts, the
implied outflow velocity would be similar to that derived from
Fe~XVII and O ions. \citet{holczer05} argued that errors in the
calculated wavelengths would be distributed randomly in both
directions, and therefore, the systematic shift observed in the
data is unlikely a result of theoretical wavelength errors.
Furthermore, it was shown that the HULLAC errors for the strongest
2p--3d transitions (those which determine the UTA centroid) of the
Fe L-shell ions are generally much smaller than 30~m\AA\
\citep{brown02, holczer05}. The present work, however, shows that
these indirect indications concerning the HULLAC accuracy for the
Fe~M-shell UTA wavelengths were misleading and that the assumption
that these wavelengths could be used for velocity measurements was
premature.

Recently, \citet{gu05} showed that a combined configuration
interaction and second order many-body perturbation theory (MBPT)
is capable of predicting wavelengths of $2\to 3$ transitions of
Fe~XVII--XXIV to within a few m{\AA}. The $2\to 3$ lines of the
Fe~M-shell UTA are similar to those of Fe XVII-XXIV, except for
the addition of M-shell spectator electrons. Following the
unexpected result of \citet{holczer05}, in this paper, we apply an
improved version of the MBPT method to the $2\to 3$ transitions of
Fe~VI--XVI. The resulting wavelengths of the UTA lines are found
to be systematically larger than HULLAC ones by 15--45~m{\AA}. We
also give improved autoionization rates, which are significantly
larger than the previous HULLAC results. The new data are used to
model the NGC~3783 spectrum, and the outflow velocity derived from
Fe~M-shell UTA is found to be consistent with that associated with
Fe~XVII and O~VII--VIII ions.

In \S\ref{theory}, we describe the detail of MBPT and its
implementation. \S\ref{result} presents the results of the calculation, and their
comparison with the previous HULLAC data set. A brief summary is given
in \S\ref{conclusion}.

\section{Theoretical Method}
\label{theory}
The present calculation is based on the Rayleigh-Schr\"odinger perturbation
theory for a multi-configurational model space. The brief
description of the method presented here closely follows \citet{lindgren74},
and readers interested in a more detailed account should consult
\citet{lindgren74}. 

The method tries to solve the Schr\"odinger equation
\begin{equation}
H_{DCB}\Psi_k = E_k\Psi_k,
\end{equation}
with perturbation expansion.
The no-pair Dirac-Coulomb-Breit (DCB) Hamiltonian $H_{DCB}$
for an $N$-electron ionic system can be written as \citep{sucher80},
\begin{equation}
H_{DCB} = \sum_i \left[h_d(i) -\frac{Z}{r_i}\right]+ \sum_{i<j}
\left(\frac{1}{r_{ij}} + B_{ij}\right),
\end{equation}
where $h_d(i)$ is the Dirac Hamiltonian for one free electron, $Z$ is the
nuclear charge, $r_i$ is the radial coordinate of the electron $i$, $r_{ij}$
is the distance between the electrons $i$ and $j$, and $B_{ij}$ is the
frequency-independent Breit interaction
\begin{equation}
B_{ij} = -\frac{1}{2r_{ij}}\left[\mathbf{\alpha_i}\cdot\mathbf{\alpha_j} + 
  \frac{(\mathbf{\alpha_i}\cdot
    \mathbf{r_{ij}})(\mathbf{\alpha_j}\cdot\mathbf{r_{ij}})}{r_{ij}^2}\right],
\end{equation}
where $\mathbf{\alpha_i}$ is a matrix vector constructed from Pauli spin
matrices $\mathbf{\sigma_i}$
\begin{equation}
\mathbf{\alpha_i} = \left(\begin{array}{cc} 0 & \mathbf{\sigma_i}
  \\ \mathbf{\sigma_i} & 0 \\ \end{array}\right).
\end{equation}

$H_{DCB}$ is split up into a model
Hamiltonian $H_0$, and a perturbation $V$. In the present case, a
convenient choice is
\begin{eqnarray}
H_0 &=& \sum_i \left[h_d(i) + U(r_i)\right] \nonumber\\
V &=& -\sum_i\left[\frac{Z}{r_i}+U(r_i)\right] + \sum_{i<j}
\left(\frac{1}{r_{ij}} + B_{ij}\right),
\end{eqnarray}
where $U(r)$ is a model potential including the screening effects of all
electrons, whose appropriate choice makes $V$ as small as possible. In many
applications, $U(r)$ is
taken to be the Hartree-Fock potential of a suitable configuration. Such a
choice simplifies the many-body expansion because many terms in the series vanish
exactly. However, Hartree-Fock potential is non-local and complicates the
generation of the zero-th order radial wavefunctions. In the present
implementation, $U(r)$ is approximated by a local central potential, and is
derived from a Dirac-Fock-Slater self-consistent
field calculation, which minimizes the weighted mean energy of all relevant
configurations. As long as the perturbation potential $V$ is kept small
enough, the exact choice for $U(r)$ does not affect the final results.

The eigenfunctions $\Phi_k$ and eigenvalues $E^0_k$ of $H_0$ are
easily obtained by forming Slater determinants from
single-electron wavefunctions once $U(r)$ is determined. A subset
of $\Phi_k$ will define a model space $M$, and the remaining
states belong to the orthogonal space $N$. A projection operator
$P$ is defined for $M$, which produces a state in the model space
when operated on an eigenfunction of the full Hamiltonian
\begin{equation}
\Psi^0_k = P\Psi_k,
\end{equation}
where $\Psi^0_k$ is generally a linear combination of the subset of $\Phi_k$
that belong to the model space $M$.

As described in \citet{lindgren74}, one can define a wave operator, $\Omega$,
that transforms $\Psi^0_k$ back to $\Psi_k$,
\begin{equation}
\Psi_k = \Omega \Psi^0_k.
\end{equation}
The original Schr\"odinger equation can then be transformed to
\begin{equation}
H_{eff}\Psi^0_k = PH\Omega\Psi^0_k = E_k\Psi^0_k.
\end{equation}
This equation defines an effective Hamiltonian in the model space
\begin{equation}
H_{eff} = PH_0P + PV\Omega,
\end{equation}

whose eigenvalues are the true eigenenergies of the full Hamiltonian.
The effective Hamiltonian is generally non-hermitian, and the
eigenfunctions, $\Psi^0_k$ are not necessarily orthogonal. It
follows from the Schr\"odinger equation and the definition of
$\Omega$ that the wave operator satisfies
\begin{equation}
\left[\Omega,H_0\right] = V\Omega - \Omega V\Omega,
\end{equation}
which is the starting point of the many-body expansion formula. The first
order expansion of $PV\Omega$ can be written as
\begin{equation}
<\Phi_i|V\Omega^{(1)}|\Phi_j> = <\Phi_i|V|\Phi_j>+\sum_{r\in N}
\frac{<\Phi_i|V|\Phi_r><\Phi_r|V|\Phi_j>}{E^0_j-E^0_r},
\end{equation}
in terms of matrix elements in the model space $M$. The first order effective
Hamiltonian is then
\begin{equation}
<\Phi_i|H^{(1)}_{eff}|\Phi_j> = H_{DCB}^{ij} + \sum_{r\in
  N}\frac{V^{ir}V^{rj}}{E^0_j-E^0_r},
\end{equation}
where $H_{DCB}^{ij}=<\Phi_i|H_{DCB}|\Phi_j>$, and $V^{ir}=<\Phi_i|V|\Phi_r>$.
By solving the generalized eigenvalue problem for the first order effective
Hamiltonian, one obtains the eigenvalues in second order.

The energy levels obtained in the present work are based on the solution of
this generalized eigenvalue problem. This is slightly different from the method
used in \citet{gu05}, where following \citet{vilkas99}, the total second order
energy is expressed as
\begin{equation}
E_k = E_k^{CI} + E_k^2,
\end{equation}
where $E_k^{CI}$ is the eigenvalues of $PH_{DCB}P$ in the model space with the
corresponding eigenvectors $b^0_{ki}$, and $E_k^2$ is
the second order correction due to correlations in the space $N$
\begin{equation}
E_k^2 = \sum_{r\in N}\sum_{i,j\in M}b^0_{ki}b^0_{kj}\frac{V^{ir}V^{rj}}{E^0_j -
  E^0_r}.
\end{equation}
This is in fact a good approximation of the generalized eigenvalue
solution for $H^{(1)}_{eff}$. If one replaces the eigenvectors $b^0_{ki}$ with the true
eigenvectors $b_{ki}$ of $H^{(1)}_{eff}$, the above expression becomes the exact
eigenvalues of the first order effective Hamiltonian. The uncertainties
introduced by
using the eigenvectors of $PH_{DCB}P$ instead of $H^{(1)}_{eff}$ are generally
of higher order, making it a valid approximation. Nonetheless, the more
accurate method of solving the generalized eigenvalue problem for
$H^{(1)}_{eff}$ is used in the present work.

In the present work, we restrict the calculation to the ground state energies
of $1s^22l^83l^q$ ($1 \le q \le 11$) configurations, and the energies for
states in the $1s^22l^73l^{q+1}$ configuration that are connected to the
ground state by
electric dipole transitions. These are the absorption lines that appear in AGN
where the ions are predominantly in the ground states. The model potential are
calculated separately for the $1s^22l^83l^q$ and $1s^22l^73l^{q+1}$
configurations. The model space $M$ consists of these configurations, and the $N$
space consists of all single and double excitations from them.
Because the $N$ space contains not only bound states, but also
continuum, a finite basis set method is used to evaluate the
perturbation expansion to convergence. The basis set is derived by imposing
the boundary condition $Q(r_b)/P(r_b) = b$ for the single electron radial
wavefunctions, where $Q(r)$ and $P(r)$ are the small and large components of
the Dirac
spinor, $r_b$ is the boundary radius, which is chosen to be large enough to
contain the $1s$, $2l$, and $3l$ wavefunction amplitudes, and $b$ is an
arbitrary constant chosen to be $\kappa/2r_bc$, where $\kappa$ is the
relativistic angular momentum quantum number of the one-electron orbital, and
$c$ is the speed of light. With
this choice, the boundary condition reduces to $P^\prime(r_b)/P(r_b) = 0$ in the
non-relativistic limit.

Several small corrections to the Hamiltonian are also included in the
calculations, namely, finite nuclear size, nuclear recoil, vacuum
polarization, and electron self-energy. These are all taken into account with
standard procedures of the atomic structure theory.

In addition to energies, we also calculate the oscillator
strengths and radiative rates for $2l$-$3l^\prime$ transitions
that are connected to the ground state of each ion. Because the
$1s^22l^73l^{q+1}$ states are highly autoionizing, the
autoionization rates dominate the total lifetime of these levels.
We calculate all possible autoionization channels between
$1s^22l^73l^{q+1}$ and $1s^22l^83l^{q-1}$ configurations. In
calculating the radiative and autoionization rates, one may either
use the eigenvectors $b^0_{ki}$ of $PH_{DCB}P$, or the generalized
eigenvectors $b_{ki}$ of $H^{(1)}_{eff}$. We have verified that
the differences in the two choices are within $\sim$ 20\% for the
ions considered in the present work, and we adopt the results of
the latter method, which is generally not very different from
that of HULLAC.

For ions from Fe~I--V, we have not carried out the MBPT calculation, since
MBPT is expected to be less accurate for near neutral ions, and also because
these ions are more complex. To give a complete database for the Fe~M-shell
UTA, we have made simple configuration interaction calculations for them, and
present the results along with MBPT calculations for Fe~VI--XVI.

\section{Results}
\label{result} 
\begin{figure}
\epsscale{0.7}
\includegraphics[width=6in]{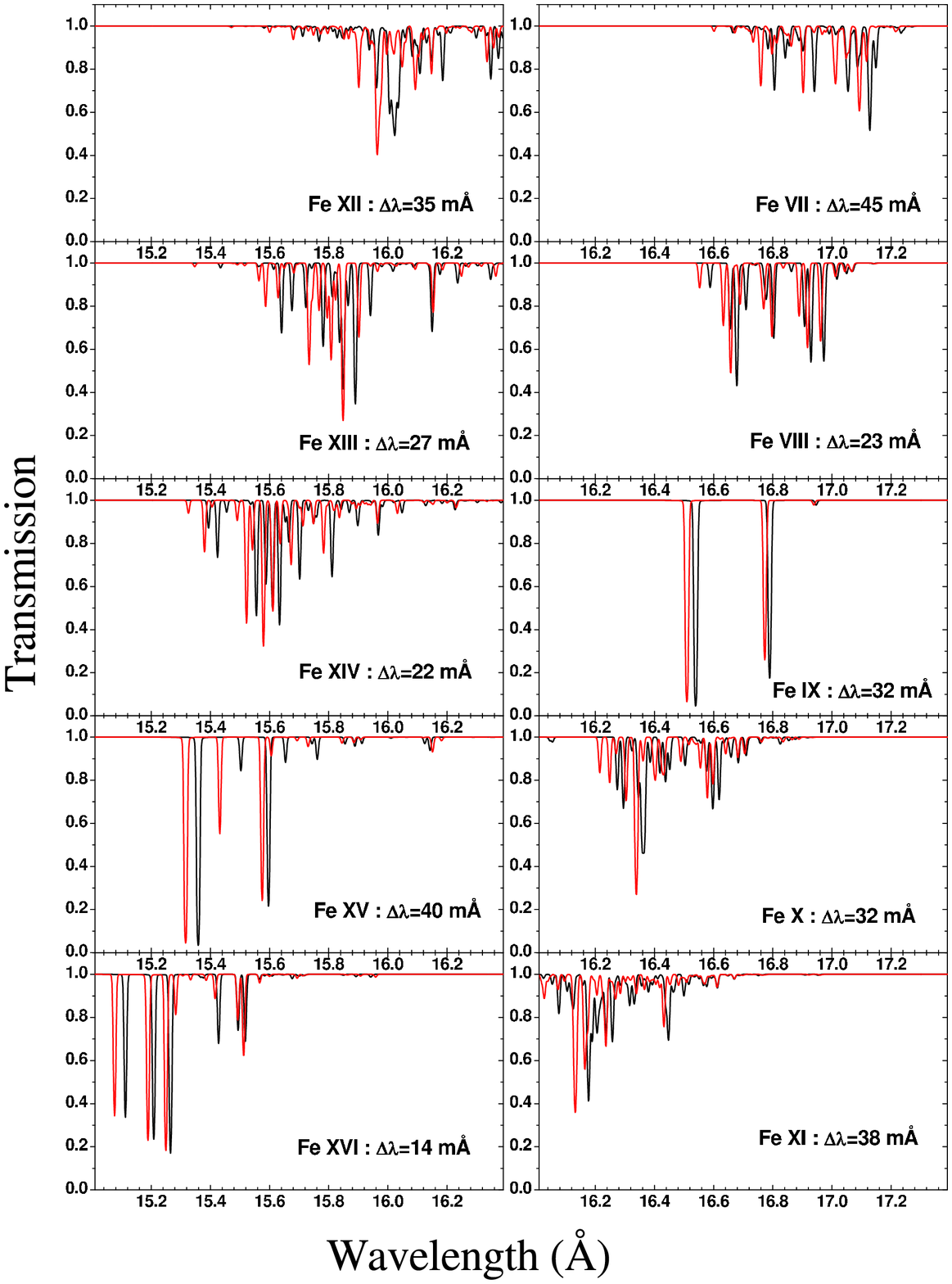}
\caption{\label{fig:flambda}  Absorption spectra of $2\to 3$
transitions from the ground state of Fe~VII--XVI at an arbitrary
column density and turbulent velocity of, respectively, 8$\times
10^{15}$~cm$^{-2}$ and 100~\kms. Black lines are the present
results and red lines are the HULLAC calculation of
\citet{behar01}. The value of $\Delta \lambda$ shown in each panel
is the difference between the oscillator strength weighted
wavelengths of the present and HULLAC calculations. }
\end{figure}

In Figure~\ref{fig:flambda}, we compare the
wavelengths and absorption oscillator strengths of the present
calculation and those of \citet{behar01} for Fe~VII--XVI. As can
be seen, the oscillator strengths in the two calculations agree
well with each other except for a few isolated weak transitions.
However, the wavelengths in the present work are systematically
larger than the HULLAC results by 15--45~m{\AA} for some of the
strongest lines. The differences between the oscillator strength
weighted wavelengths in the two calculations are also shown in the
figure. We have no simple explanation for why the MBPT wavelengths
are systematically longer than the configuration-interaction
method of HULLAC. 

Table~\ref{tab:lines} gives the complete list of Fe~M-shell UTA transitions
with the ground states as the lower levels. The database is a combination of
MBPT results for Fe~VI--XVI and configuration interaction results for
Fe~I--V. The tabulated data consist of wavelengths, oscillator strengths,
radiative transition rates, and autoionization rates. The autoionization rates
are sums over all relevant autoionization channels. The radiative
transition rates are those between the ground states and the specific upper
levels, i.e., possible radiative decays to excited states are not included. The
omission of these alternative radiative channels should have
negligible effects on the total natural widths of the upper levels, since
autoionization rates dominate the radiative rates in most cases, and for a few
lines of Fe~XVI and XV, where radiative rates are comparable to
autoionization rates, decays to the ground states are the dominant branches.

In addition to wavelengths, another major difference between the
present and the earlier calculations of \citet{behar01} is in the
autoionization rates. The present autoionization rates are
significantly larger than the previous results, especially for
lower charge states. The discrepancy is caused by additional
autoionization channels that are included here. Let us take the
$1s^22s^22p^53s^23p^63d$ excited levels of Fe~IX as an example. In
\citet{behar01}, only the autoionization channels that involve the
ejection of the 3d electron, i.e. toward $1s^22s^22p^6(3s3p)^5$,
were considered. The present addition of channels to levels of
$1s^22s^22p^6(3s3p)^63d$ gives total autoionization rates that are
$\sim$3.7 times higher. The increased autoionization rates broaden
further the natural line width and may have implications for the
curve of growth if turbulent broadening is particularly small. On
the other hand, if the source has significant turbulent broadening
exceeding hundreds of km\,s$^{-1}$, as is the usual case for AGN
outflows, this effect is unlikely to be observed.

\begin{figure}
\epsscale{0.7}
\includegraphics[width=5in, angle=90]{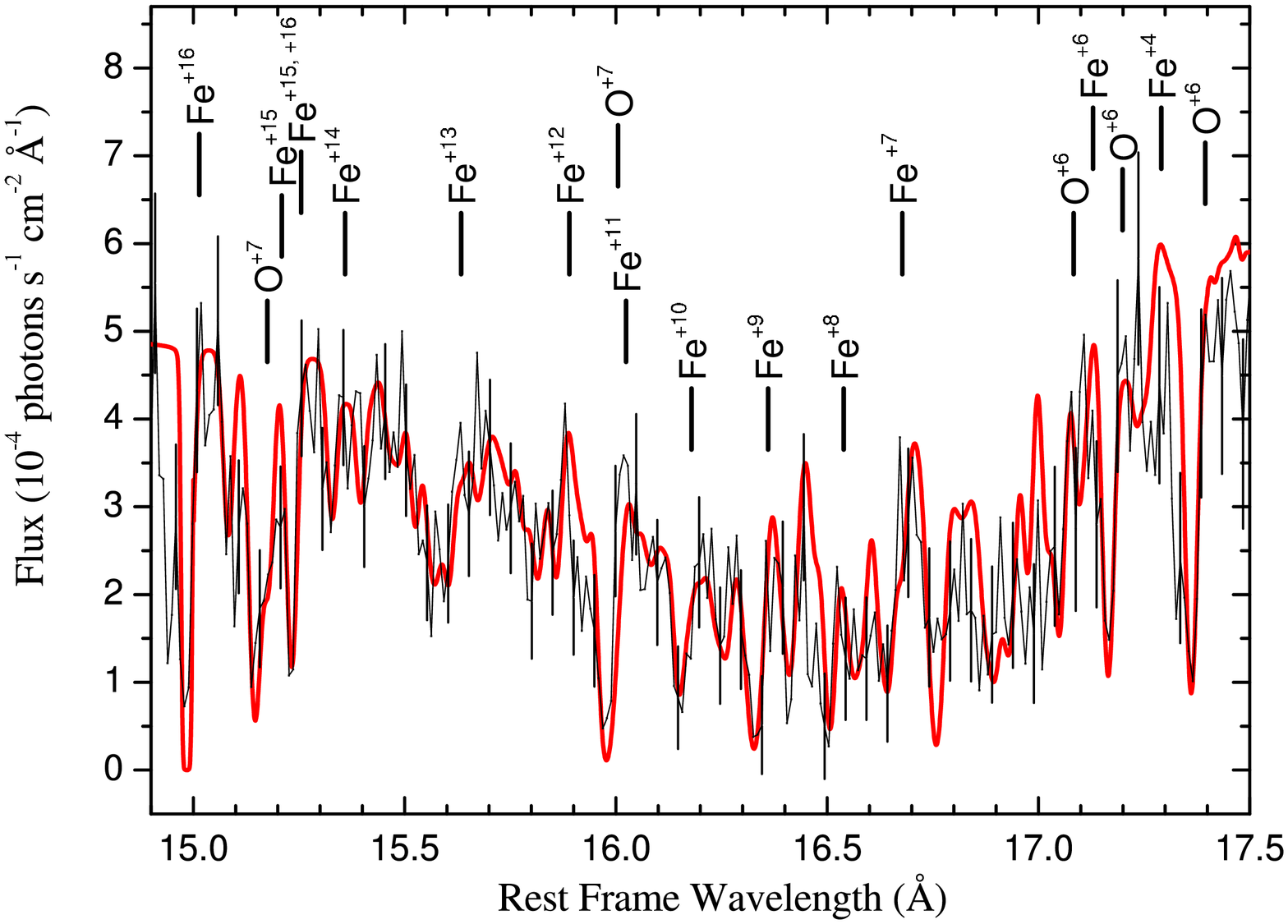}
\caption{\label{fig:ngc3783} \textit{Chandra} HETGS spectra of NGC~3783 in the
  Fe~M-shell UTA region. Black line with error bars (plotted every six data
  points) are the data, red line is
  the model with a uniform outflow velocity of $-590$~km\,s$^{-1}$ using the
  present atomic data for Fe~VI--XVI.}
\end{figure}

\citet{holczer05} recently reanalyzed the HETGS spectra of
NGC~3783 in the Fe M-shell UTA region. The HULLAC data of
\citet{behar01} was used in that analysis. The major conclusion
was that the outflow velocity of the gas associated with
Fe~M-shell ions are significantly less than that derived from the
Fe~XVII, O~VII, and O~VIII lines. By comparing the deepest
observed centroid of each ion and the HULLAC-calculated
individual-ion centroid for Fe~IX--XVI, velocities between $-374$
and $118$~km\,s$^{-1}$ were derived. The same comparison, but now
carried out with the present MBPT data is shown in
Table~\ref{tab:ngc3783}. Except for a few ions where the
identification of a unique UTA centroid becomes difficult, the
velocities range between $-450$ and $-780$~\kms. We have fitted a
uniform outflow velocity ($-590$~km\,s$^{-1}$) model to the same
NGC~3783 spectrum of \citet{holczer05}, using the present atomic
data for Fe~VI--XVI. The results are shown in
Figure~\ref{fig:ngc3783}. Inspection of Figure~\ref{fig:ngc3783}
shows that no systematic discrepancies between the observed and
theoretical absorption features for the Fe~M-shell UTA can be
readily identified, and therefore, the Fe~M-shell and other more
ionized ions appear to belong to the same kinematic system.

We have been able to identify one single laboratory measurement of
Fe~XVI lines against which we may check the accuracy of the
present calculation. \citet{brown01} gave the wavelengths of the
three strongest $2p$--$3d$ transitions identified in the emission
spectrum taken at the Lawrence Livermore National Laboratory's
electron beam ion trap. One of the three lines is blended with a
strong Fe~XVII line, and the other two lines are unblended, which
allowed the accurate determination of their wavelengths. The two
lines are
$2p^63s$($J=1/2$)--$2p_{1/2}2p_{3/2}^43s3d_{3/2}$($J=1/2,3/2$)
transitions, and have measured wavelengths of 15.208(4) and
15.115(6)~{\AA}. The present work gives wavelengths of 15.2086 and
15.1127~{\AA} respectively, which agree very well with the
measured values. In the contrary, \citet{behar01} gave wavelengths
of 15.1890 and 15.0762~{\AA} respectively, which are 20-40~m{\AA}
too small. \citet{brown01} also identified the strongest Fe~XV
line, $2p^63s^2$($J=0$)--$2p_{1/2}2p_{3/2}^43s^23d_{3/2}$($J=1$),
in their spectrum, but did not give its measured wavelength.
Through inspection of Figure~1 in their paper and private
communications, we estimate that its wavelength is
15.353(5)~{\AA}, which also agrees reasonably well with the
present result of 15.3588~{\AA}, but 37~m{\AA} larger than the
HULLAC result. Unfortunately, the Fe~XV and XVI absorption lines
are relatively weak in the NGC~3783 spectrum, and Fe~XVI lines are
severely blended with Fe~XVII lines. Nevertheless, the excellent
agreement between the present results and the measurement for
these few lines lends credibility to the MBPT treatment of the
entire iso-nuclear sequence.

\section{Conclusions}
\label{conclusion} We have developed a second order many-body
perturbation theory with multi-reference model space. We
apply the method to the calculation of wavelengths, oscillator
strengths, and autoionization rates of Fe~M-shell UTA arising from
$2p$--$3d$ transitions originating from the ground state of Fe
VI-XVI. The wavelengths obtained in the present work are
systematically larger than the HULLAC calculation of
\citet{behar01}; the present oscillator strengths of the strong
absorption lines agree well with the HULLAC results; and the
autoionization rates of \citet{behar01} are found to be missing
important autoionization channels, especially for lower charge
states. Using the present data for Fe~VI--XVI, we find no evidence
that the outflow velocity of the gas associated with Fe~M-shell
ions is different from that derived from Fe~XVII and O~VII--VIII
ions, as claimed in the recent analysis of \citet{holczer05} using
the HULLAC data. A complete list of Fe~M-shell UTA lines are
given, which include the present MBPT calculation for Fe~VI--XVI,
and a simple configuration interaction calculation for Fe~I--V.
We recommend that this
new database should be preferred over the earlier HULLAC
calculation of \citet{behar01} in the analyses of future
absorption spectroscopy where Fe~M-shell UTA is prominent.

M.F. Gu and S.M. Kahn acknowledges the partial support of NASA grants
NAG5-5419 and NNGG04GL76G. The research at the Technion was supported by The
Israel Science Foundation (grant no. 28/03). E. Behar thanks the Stanford
group for their hospitality during a visit in July 2005. We thank Shai Kaspi
for making the 900~ks HETGS integrated spectrum available to us.


\clearpage
\begin{deluxetable}{ *{7}{c}l }
\tabletypesize{\scriptsize}
\tablecaption{\label{tab:lines}Wavelengths, oscillator strengths, radiative
  decay rates, and autoionization rates of $2\to 3$ transitions from the
  ground states of Fe~I--XVI. For Fe~VI-XVI, the MBPT method is used and for
  Fe~I--VI, the standard configuration interaction method is used.}
\tablehead{
\colhead{Ion} &
\colhead{Upper} &
\colhead{$\lambda$ ({\AA})} &
\colhead{$A^r$ (s$^{-1}$)} &
\colhead{$A^a$ (s$^{-1}$)} &
\colhead{$A^a$ (s$^{-1}$)\tablenotemark{a}} &
\colhead{$f_{ij}$} &
\colhead{Configuration ($J$)\tablenotemark{b}}
}
\startdata 
 XVI &   25 & 15.2652 & $ 1.41$E$+13$ & $ 2.11$E$+13$ & $ 2.63$E$+13$ & $ 9.88$E$-01$ & $2p_{1/2}3s_{1/2}3d_{5/2}$($J=\frac{3}{2}$)\\
 XVI &   26 & 15.2086 & $ 2.38$E$+13$ & $ 1.68$E$+10$ & $ 1.39$E$+13$ & $ 8.27$E$-01$ & $2p_{1/2}3s_{1/2}3d_{3/2}$($J=\frac{1}{2}$)\\
 XVI &   27 & 15.1127 & $ 8.92$E$+12$ & $ 4.03$E$+13$ & $ 6.33$E$+13$ & $ 6.11$E$-01$ & $2p_{1/2}3s_{1/2}3d_{3/2}$($J=\frac{3}{2}$)\\
  XV &   16 & 15.3588 & $ 1.78$E$+13$ & $ 3.76$E$+13$ & $ 4.27$E$+13$ & $ 1.89$E$+00$ & $2p_{1/2}3d_{3/2}$($J=1$)\\
  XV &   14 & 15.5965 & $ 7.64$E$+12$ & $ 4.57$E$+13$ & $ 6.93$E$+13$ & $ 8.36$E$-01$ & $2p_{3/2}^{3}3d_{5/2}$($J=1$)\\
  XV &   18 & 14.1828 & $ 3.06$E$+12$ & $ 1.10$E$+14$ & $ 3.24$E$+14$ & $ 2.77$E$-01$ & $2s_{1/2}3p_{3/2}$($J=1$)\\
 XIV &   51 & 15.6339 & $ 6.11$E$+12$ & $ 4.51$E$+13$ & $ 3.58$E$+13$ & $ 4.48$E$-01$ & $2p_{1/2}3p_{1/2}3d_{3/2}$($J=\frac{3}{2}$)\\
 XIV &   55 & 15.5555 & $ 1.16$E$+13$ & $ 9.90$E$+13$ & $ 8.53$E$+13$ & $ 4.22$E$-01$ & $2p_{1/2}3p_{1/2}3d_{3/2}$($J=\frac{1}{2}$)\\
 XIV &   53 & 15.6098 & $ 5.06$E$+12$ & $ 4.88$E$+13$ & $ 9.43$E$+13$ & $ 3.70$E$-01$ & $2p_{1/2}3p_{3/2}3d_{3/2}$($J=\frac{3}{2}$)\\
XIII &   31 & 15.8898 & $ 4.87$E$+12$ & $ 1.76$E$+14$ & $ 2.14$E$+14$ & $ 5.53$E$-01$ & $2p_{3/2}^{3}3p_{1/2}3p_{3/2}3d_{5/2}$($J=1$)\\
XIII &   33 & 15.8495 & $ 4.13$E$+12$ & $ 1.01$E$+14$ & $ 1.30$E$+14$ & $ 4.67$E$-01$ & $2p_{3/2}^{3}3d_{5/2}$($J=1$)\\
XIII &   38 & 15.7811 & $ 2.31$E$+12$ & $ 2.14$E$+14$ & $ 1.32$E$+14$ & $ 2.59$E$-01$ & $2p_{1/2}3p_{1/2}3p_{3/2}3d_{3/2}$($J=1$)\\
 XII &   95 & 16.0253 & $ 4.06$E$+12$ & $ 1.19$E$+14$ & $ 1.26$E$+14$ & $ 2.35$E$-01$ & $2p_{3/2}^{3}3p_{1/2}3p_{3/2}^{2}3d_{5/2}$($J=\frac{5}{2}$)\\
 XII &   97 & 16.0151 & $ 4.77$E$+12$ & $ 5.39$E$+14$ & $ 1.04$E$+14$ & $ 1.84$E$-01$ & $2p_{1/2}3p_{3/2}3d_{3/2}$($J=\frac{3}{2}$)\\
 XII &   93 & 16.0345 & $ 9.30$E$+12$ & $ 2.08$E$+14$ & $ 1.01$E$+14$ & $ 1.79$E$-01$ & $2p_{3/2}^{3}3p_{3/2}^{3}3d_{5/2}$($J=\frac{1}{2}$)\\
  XI &   90 & 16.1761 & $ 5.21$E$+12$ & $ 3.08$E$+14$ & $ 1.00$E$+14$ & $ 2.86$E$-01$ & $2p_{1/2}3p_{3/2}^{2}3d_{3/2}$($J=3$)\\
  XI &   89 & 16.1788 & $ 8.63$E$+12$ & $ 2.98$E$+14$ & $ 9.94$E$+13$ & $ 2.03$E$-01$ & $2p_{3/2}^{3}3p_{1/2}3p_{3/2}^{3}3d_{5/2}$($J=1$)\\
  XI &   74 & 16.2578 & $ 3.36$E$+12$ & $ 3.07$E$+14$ & $ 1.35$E$+14$ & $ 1.86$E$-01$ & $2p_{1/2}3p_{1/2}3p_{3/2}^{3}3d_{3/2}$($J=3$)\\
   X &   39 & 16.3581 & $ 8.48$E$+12$ & $ 3.75$E$+14$ & $ 1.42$E$+14$ & $ 3.40$E$-01$ & $2p_{1/2}3p_{3/2}^{3}3d_{3/2}$($J=\frac{3}{2}$)\\
   X &   38 & 16.3661 & $ 5.59$E$+12$ & $ 4.39$E$+14$ & $ 1.90$E$+14$ & $ 3.37$E$-01$ & $2p_{1/2}3p_{3/2}^{3}3d_{3/2}$($J=\frac{5}{2}$)\\
   X &   43 & 16.2951 & $ 3.47$E$+12$ & $ 7.50$E$+14$ & $ 1.99$E$+14$ & $ 2.07$E$-01$ & $2p_{3/2}^{3}3p_{3/2}^{3}3d_{5/2}$($J=\frac{5}{2}$)\\
  IX &    3 & 16.5392 & $ 1.29$E$+13$ & $ 6.31$E$+14$ & $ 1.72$E$+14$ & $ 1.59$E$+00$ & $2p_{1/2}3d_{3/2}$($J=1$)\\
  IX &    2 & 16.7894 & $ 6.93$E$+12$ & $ 5.85$E$+14$ & $ 9.61$E$+13$ & $ 8.79$E$-01$ & $2p_{3/2}^{3}3d_{5/2}$($J=1$)\\
  IX &    1 & 16.9464 & $ 8.44$E$+10$ & $ 5.57$E$+14$ & $ 6.26$E$+13$ & $ 1.09$E$-02$ & $2p_{3/2}^{3}3d_{3/2}$($J=1$)\\
VIII &   27 & 16.6784 & $ 1.03$E$+13$ & $ 6.37$E$+14$ & $ 2.16$E$+14$ & $ 4.30$E$-01$ & $2p_{1/2}3d_{3/2}^{2}$($J=\frac{3}{2}$)\\
VIII &   16 & 16.9289 & $ 4.72$E$+12$ & $ 6.42$E$+14$ & $ 1.97$E$+14$ & $ 3.04$E$-01$ & $2p_{3/2}^{3}3d_{3/2}3d_{5/2}$($J=\frac{5}{2}$)\\
VIII &   19 & 16.8030 & $ 3.38$E$+12$ & $ 6.01$E$+14$ & $ 1.64$E$+14$ & $ 2.14$E$-01$ & $2p_{1/2}3d_{3/2}3d_{5/2}$($J=\frac{5}{2}$)\\
 VII &   46 & 16.9404 & $ 4.03$E$+12$ & $ 6.26$E$+14$ & $ 2.42$E$+14$ & $ 1.73$E$-01$ & $2p_{1/2}3d_{3/2}^{2}3d_{5/2}$($J=2$)\\
 VII &   26 & 17.1276 & $ 2.52$E$+12$ & $ 7.69$E$+14$ & $ 3.74$E$+14$ & $ 1.55$E$-01$ & $2p_{3/2}^{3}3d_{3/2}3d_{5/2}^{2}$($J=3$)\\
 VII &   25 & 17.1289 & $ 3.45$E$+12$ & $ 6.13$E$+14$ & $ 2.90$E$+14$ & $ 1.52$E$-01$ & $2p_{3/2}^{3}3d_{3/2}^{2}3d_{5/2}$($J=2$)\\
  VI &   24 & 17.2656 & $ 3.80$E$+12$ & $ 7.30$E$+14$ & $ 3.88$E$+14$ & $ 2.55$E$-01$ & $2p_{3/2}^{3}3d_{3/2}^{2}3d_{5/2}^{2}$($J=\frac{5}{2}$)\\
  VI &   35 & 17.2105 & $ 3.31$E$+12$ & $ 5.93$E$+14$ & $ 3.13$E$+14$ & $ 1.47$E$-01$ & $2p_{3/2}^{3}3d_{3/2}3d_{5/2}^{3}$($J=\frac{3}{2}$)\\
  VI &   39 & 17.1881 & $ 2.72$E$+12$ & $ 7.06$E$+14$ & $ 2.33$E$+14$ & $ 1.21$E$-01$ & $2p_{3/2}^{3}3d_{3/2}^{2}3d_{5/2}^{2}$($J=\frac{3}{2}$)\\
\enddata

\footnotesize
\tablenotetext{a}{Total autoionization rates in the HULLAC calculation of
  \citet{behar01}. This column is not present in the electronic version of
  this table, and is only included here to illustrated the difference between
  the present and HULLAC results.}

\tablenotetext{b}{The configurations are given in the $jj$ coupling notation, and
  closed subshells are omitted to save space.}

\tablecomments{Only three strongest absorption lines per ion from Fe~VI--XVI
  are shown here. The entire table is included in the electronic version of
  the paper as a machine readable file.}
\end{deluxetable}

\clearpage
\begin{deluxetable}{lcccccl}
\tabletypesize{\scriptsize}
\tablecolumns{5} \tablewidth{0pt}
\tablecaption{Best-fit velocities and column densities for ions
detected in the 14.9--17.5~\AA\ region of the \hetgs\ spectrum of
\ngc 3783. \label{tab:ngc3783}}
\tablehead{
   \colhead{Ion} &
   \colhead{$\lambda _{\mathrm {Observed}}$} &
   \colhead{$\lambda _{\mathrm {Rest}}$\tablenotemark{a}} &
   \colhead{$\lambda _{\mathrm {Model}}$\tablenotemark{b}} &
   \colhead{Outflow Velocity \tablenotemark{c}} &
   \colhead{Ion Column Density } &
   \colhead{Configuration ($J$)}\\
   \colhead{} &
   \colhead{(\AA)} &
   \colhead{(\AA)} &
   \colhead{(\AA)} &
   \colhead{(\kms)} &
   \colhead{(10$^{16}$~cm$^{-2}$)} &
   \colhead{}
}
  \startdata
O$^{+7}$ & 15.144 $\pm$ 0.003 \tablenotemark{d} & 15.176 & 15.146 & $-$631 $\pm$
59 & 400 $\pm$ 60 & $4p$($J=1/2,3/2$)\\
         & 15.970 $\pm$ 0.005 \tablenotemark{e} & 16.006 & 15.977 & $-$675 $\pm$
94 &  & $3p$($J=1/2,3/2$)\\
O$^{+6}$ & 17.351 $\pm$ 0.005 & 17.395 & 17.361 & $-$759 $\pm$ 86 & 110 $\pm$ 20
 & $1s5p$($J=1$)\\
         & 17.161 $\pm$ 0.005 & 17.199 & 17.165 & $-$663 $\pm$ 87 & & $1s6p$($J=1$)  \\
         & 17.048 $\pm$ 0.005 & 17.084 & 17.050 & $-$632$ \pm$ 88  &  & $1s7p$($J=1$)\\
Fe$^{+16}$ & 14.980 $\pm$~0.003 & 15.013 & 14.985 & $-$659 $\pm$ 60 & 3.0 $\pm$
0.5 & $2p_{1/2}3d_{3/2}$($J=1$) \\
           & 15.231 $\pm$~0.002 \tablenotemark{f} & 15.261 & 15.232 & $-$590
$\pm$ 39 & & $2p_{3/2}3d_{5/2}$($J=1$)\\
\hline
Fe$^{+15}$ & 15.231 $\pm$~0.002 \tablenotemark{f} & 15.265 & 15.232 & $-$668 $\pm$ 39 & 1.6 $\pm$ 0.4  & $2p_{1/2}3s_{1/2}3d_{5/2}$($J=\frac{3}{2}$)\\
Fe$^{+14}$ & 15.322 $\pm$ 0.006 & 15.359 & 15.329 & $-$723 $\pm$ 117 & 0.4 $\pm$ 0.1 & $2p_{1/2}3d_{3/2}$($J=1$) \\
Fe$^{+13} $ & 15.569 $\pm$ 0.008 \tablenotemark{g} & 15.634 & 15.603 & $-$1247 $\pm$ 154 & 1.1 $\pm$ 0.4  & $2p_{1/2}3p_{1/2}3d_{3/2}$($J=\frac{3}{2}$) \\
Fe$^{+12} $ & 15.844 $\pm$ 0.014 \tablenotemark{g} & 15.890 & 15.816 & $-$868 $\pm$ 264 & 1.2 $\pm$ 0.3  & $2p_{3/2}^{3}3p_{1/2}3p_{3/2}3d_{5/2}$($J=1$) \\
Fe$^{+11}$ & 15.970 $\pm$ 0.005 \tablenotemark{e} & 16.024 & 15.977 & $-$1011 $\pm$ 94 & 2.3 $\pm$ 1.5  & $2p_{3/2}^{3}3p_{1/2}3p_{3/2}^{2}3d_{5/2}$($J=\frac{5}{2}$)\\
Fe$^{+10}$ & 16.154 $\pm$ 0.005 & 16.179 & 16.150 & $-$464 $\pm$ 93 & 3.0 $\pm$ 1.2 & $2p_{1/2}3p_{3/2}^{2}3d_{3/2}$($J=3$) \\
Fe$^{+9}$ & 16.329 $\pm$ 0.005 & 16.360 & 16.327 & $-$568 $\pm$ 92 & 5.5 $\pm$ 1.0  & $2p_{1/2}3p_{3/2}^{3}3d_{3/2}$($J=\frac{3}{2}$) \\
Fe$^{+8}$ & 16.496 $\pm$ 0.004 & 16.539 & 16.507 & $-$780 $\pm$ 73 & 4.0 $\pm$ 0.5  & $2p_{1/2}3d_{3/2}$($J=1$)\\
Fe$^{+7}$ \tablenotemark{h} & \nodata & 16.678 & \nodata & \nodata & 3.0 $\pm$ 1.0 & $2p_{1/2}3d_{3/2}^{2}$($J=\frac{3}{2}$) \\
Fe$^{+6}$ \tablenotemark{h} & \nodata & 17.129 & \nodata & \nodata & 2.0 $\pm$ 0.7 & $2p_{1/2}3d_{3/2}^{2}3d_{5/2}$($J=2$) \\
Fe$^{+5}$ \tablenotemark{h} & \nodata & 17.266 & \nodata & \nodata & 1.5 $\pm$ 0.6 & $2p_{3/2}^{3}3d_{3/2}^{2}3d_{5/2}^{2}$($J=\frac{5}{2}$) \\
Fe$^{+4}$ \tablenotemark{h} & \nodata & 17.328 & \nodata & \nodata & 1.5 $\pm$ 0.6  & $2p_{3/2}^{3}3d_{3/2}^{2}3d_{5/2}^{3}$($J=1$)\\
Fe$^{+3}$ \tablenotemark{h} & \nodata & 17.387 & \nodata & \nodata & $\leq$ 0.8  & $2p_{3/2}^{3}3d_{3/2}^{2}3d_{5/2}^{4}$($J=\frac{3}{2}$)\\
%
\enddata
\footnotesize \tablenotetext{a}{ For Fe-M ions, these are
centroids of the deepest feature in each ionic spectrum. The configuration
labels given in the last column also correspond to these deepest features.}

\tablenotetext{b} { Centroid in full multi-ion model (Fig.~2).}

\tablenotetext{c} { Estimated by ($\lambda _{\mathrm Observed} -
\lambda _{\mathrm Rest}) c/ \lambda _{\mathrm Rest}$. Errors
reflect 90\% confidence intervals.}

\tablenotetext{d} { Blend of O$^{+7}$ and Fe$^{+15}$ at 15.19~\AA\
fitted for two distinct features.}

\tablenotetext{e}{ Unresolved blend of O$^{+7}$ and Fe$^{+11}$
reflected in the high velocity and large error on the Fe$^{+11}$ column
density.}

\tablenotetext{f}{ Unresolved blend of Fe$^{+16}$ and Fe$^{+15}$.}

\tablenotetext{g}{ These UTAs have multiple minima making centroid determination
difficult.
(See Fig. 2.)}

\tablenotetext{h}{ Indirect identification based on best-fit
model; Strongly blended with the high-$n$ lines of O$^{+6}$.}
\end{deluxetable}

\clearpage

\clearpage
\end{document}